\documentclass[aps,prl,showpacs,twocolumn,superscriptaddress]{revtex4}
\usepackage{graphicx,amsmath,amsfonts}

\RequirePackage{xspace}

\newcommand{\BABARPubYear}     {05}
\newcommand{\BABARPubNumber}  {029}
\newcommand{\SLACPubNumber} {11320}
\newcommand{\LANLNumber}  {0506081}

\input pubboard/babarsym.tex

\def\BaBar{\mbox{\slshape B\kern-0.1em{\smaller A}\kern-0.1em
    B\kern-0.1em{\smaller A\kern-0.2em R}}\xspace}
\def\babar{\BaBar}

\newcommand{\gevccSQ}{\ensuremath{{\mathrm{\,Ge\kern -0.1em V^2\!/}c^4}}\xspace}
\newcommand{\evcc}{\ensuremath{{\mathrm{\,e\kern -0.1em V\!/}c^2}}\xspace}

\def\Xstat{X(3872)}
\def\Ystat{Y(4260)}
\def\jsi{\jpsi}
\def\psip{\psitwos}
\def\ee{e^+e^-}
\def\mm{\mu^+\mu^-}
\def\ll{\ell^+\ell^-}
\def\pipi{\pi^+\pi^-}
\def\KK{K^+K^-}
\def\KPi{K^\pm \pi^\mp}
\def\DDbar{\ensuremath{D\overline D}\xspace}

\def\BR{{\cal B}}

\def\mrec{m_{Rec}}

\def\jee{\jsi\to\ee}
\def\jmm{\jsi\to\mm}
\def\pipiJ{\pipi\jsi}

\def\smallISR{\scalebox{0.5}{ISR}}
\def\gISR{\gamma_{\smallISR}\xspace}

\graphicspath{{./eps/}}

\begin{document}

\preprint{\babar-PUB-\BABARPubYear/\BABARPubNumber}
\preprint{SLAC-PUB-\SLACPubNumber}

\begin{flushleft}
  \babar-PUB-\BABARPubYear/\BABARPubNumber \\
  SLAC-PUB-\SLACPubNumber \\
  hep-ex/\LANLNumber \\
\end{flushleft}

\title{ \large \bf \boldmath
  Observation of a Broad Structure in the $\pipiJ$ Mass Spectrum 
  around 4.26\gevcc
}

%
\author{B.~Aubert}
\author{R.~Barate}
\author{D.~Boutigny}
\author{F.~Couderc}
\author{Y.~Karyotakis}
\author{J.~P.~Lees}
\author{V.~Poireau}
\author{V.~Tisserand}
\author{A.~Zghiche}
\affiliation{Laboratoire de Physique des Particules, F-74941 Annecy-le-Vieux, France }
\author{E.~Grauges}
\affiliation{IFAE, Universitat Autonoma de Barcelona, E-08193 Bellaterra, Barcelona, Spain }
% ...
\collaboration{The \babar\ Collaboration}
\noaffiliation

\date{\today}

\begin{abstract}
       We study initial-state radiation events, $\ee\to\gISR\pipiJ$, 
       with data collected with the \BaBar detector.  
       We observe an accumulation of events near 4.26\gevcc 
       in the invariant-mass spectrum of $\pipiJ$. 
       Fits to the mass spectrum
       indicate that a broad resonance with a mass of about 4.26\gevcc
       is required to describe the observed structure. 
       The presence of additional narrow resonances cannot be excluded.
       The fitted width of the broad resonance is $50$ to $90$\mevcc, 
       depending on the fit hypothesis.

\end{abstract}

\pacs{%
      14.40.Gx         
      13.66.Bc,        
      13.25.Gv         
     }

\maketitle

Recent observations of the $\Xstat$, 
decaying into $\pipiJ$~\cite{x-belle,cdf-B-X,d0-B-X,babar-B-X}, 
and the $Y(3940)$, decaying into $\omega\,\jsi$~\cite{y3940-belle}, 
have renewed experimental interest in charmonium spectroscopy.
We have previously reported a search for direct  $\Xstat$ production in 
$\ee$ annihilation through initial-state radiation (ISR):
$\ee\to \gISR X$~\cite{ct:BaBar-ISR_X3872}.
No signal is observed, suggesting that the $\Xstat$ is not a $1^{--}$ state, 
just as expected for a narrow state well above the $\DDbar$ threshold. 
In this Letter we present a study of the $\ee\to\gISR\pipiJ$ 
process across the charmonium mass range.

We use data collected with the \BaBar detector~\cite{babar-detector} at the 
SLAC PEP-II asymmetric-energy $\ee$ storage ring. 
These data represent an integrated luminosity of 211\invfb 
collected at $\sqrt{s}=10.58\gev$, 
near the peak of the $\Upsilon(4S)$ resonance, 
plus 22\invfb collected approximately 40\mev below this energy.

Charged-particle momenta are measured in a tracking system consisting 
of a five-layer double-sided silicon vertex tracker (SVT) and a 
40-layer central drift chamber (DCH), both situated in a 1.5-T axial 
magnetic field. 
An internally reflecting ring-imaging Cherenkov detector (DIRC) 
with quartz bar radiators 
provides charged-particle identification. A CsI electromagnetic 
calorimeter (EMC) is used to detect and identify photons and 
electrons, while muons are identified in the instrumented magnetic 
flux return system (IFR).

Electron candidates are identified by the ratio of the shower energy 
deposited in the EMC to the momentum, the shower shape, the specific 
ionization in the DCH, and the Cherenkov angle measured by the DIRC. 
Muons are identified by the depth of penetration into the IFR, the IFR cluster
geometry, and the energy deposited in the EMC.
Pion candidates are selected based on a likelihood calculated from the 
specific ionization in the DCH and SVT, and the Cherenkov angle measured
in the DIRC.
Photon candidates are identified with clusters in the EMC that have a shape 
consistent with an electromagnetic shower but without an associated 
charged track.

A candidate $\jsi$ meson is reconstructed via its decay to $\ee$ or $\mm$.
The lepton tracks must be well reconstructed, and at least one must be
identified as an electron or a muon. 
An algorithm  to associate and combine the 
energy from bremsstrahlung photons with nearby electron tracks 
is used when forming a $\jee$ candidate. 
An $\ee$ ($\mm$) pair with an invariant 
mass within $^{+33}_{-95}$ ($^{+33}_{-40}$)\mevcc of the nominal $\jsi$
mass is taken as a $\jsi$ candidate and is combined with a
pair of oppositely charged tracks that are identified as pions. 

Following an observation of an enhancement in the $\pipiJ$ mass spectrum 
during an earlier search for ISR $\Xstat$ production in a 124\invfb 
subsample of the available data, 
we chose to exclude the mass region from 4.2 to 4.4\gevcc from consideration 
during optimization of the selection criteria with the full sample 
to avoid the introduction of statistical or other biases 
in the analysis of this region.
Radiative production of the $\psip$ serves as a clean benchmark 
process~\cite{lou-psi2s} for a data-driven optimization. 
Selection criteria are chosen to maximize
$N/(3/2+\sqrt{B})$~\cite{optimization}, 
where $N$ is the total number of $\gISR\psip,\psip\to \pipiJ$ 
candidates in the 20\mevcc $\pipiJ$ mass range that
brackets the $\psip$ mass, and 
$B$ is the number of events in the $\pipiJ$ mass regions 
[3.8, 4.2]\gevcc and [4.4, 4.8]\gevcc, 
scaled to the width of the originally observed peak. 
Simulated ISR events are validated with 
the $\psip$ data and are used to extrapolate 
the selection criteria to the excluded mass region as 
appropriate for small kinematic differences due to the higher mass.

Radiative $\ee\to\gISR\pipiJ$ events are characterized 
by a small mass recoiling against the $\pipiJ$ system 
and by low missing transverse momentum. 
These properties are reflected in (1), (2), and (3) of 
the selection criteria: 
(1) there must be no additional well-reconstructed charged tracks in the event;
(2) the transverse component of the visible momentum 
    in the $\ee$ center-of-mass
    frame, including the ISR photon when it is reconstructed, 
    must be less than 2.5\gevc;
(3) the inferred value of the square of the mass recoiling 
    against the $\pipiJ$ combination ($\mrec^2$) must be within
    $[-1.02,+3.27]$\gevccSQ for $\jee$ candidates and 
    $[-1.06,+1.25]$\gevccSQ for $\jmm$ candidates;
(4) $\cos\theta_\ell$,
    where $\theta_\ell$ is the angle between the $\ell^+$
    momentum in the $\jsi$ rest frame and the $\jsi$ momentum 
    in the $\ee$ center-of-mass frame, 
    must satisfy $|\cos\theta_\ell|<0.90$.
In addition,
(5) for the $\ee$ mode, $\cos\theta_{\pi}$,
    where $\theta_{\pi}$ is the angle between the $\pi^-$
    momentum and the $\jsi$ momentum in the $\pi^+\pi^-$ rest frame,
    is required to be less than 0.90
    to reject background from misidentified low momentum
    $e^-$ in the forward region of the detector.
We do not require the ISR photon to be detected in the EMC since 
it is produced preferentially along the beam direction.

Candidate $\pipi\ll$ tracks are refitted, constrained 
to a common vertex, while the lepton pair 
is kinematically constrained to the $\jsi$ mass.
The  resulting $\pipiJ$ mass-resolution function is well-described by 
a Cauchy distribution~\cite{ct:Cauchy} 
with a full width at half maximum of 4.2\mevcc for the 
$\psip$  and 5.3\mevcc at 4.3\gevcc. 

\begin{figure}[t]
 \centering
   \includegraphics[width=8.8cm]{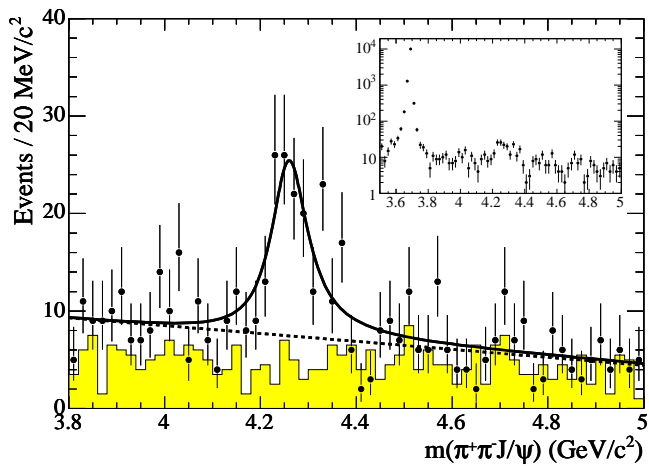}
\caption{
  The $\pipiJ$ invariant mass spectrum in the range 3.8$-$5.0\gevcc
  and (inset) over a wider range that includes the $\psip$.
  The points with error bars represent the selected data and the shaded
  histogram represents the scaled data from neighboring $\ee$ and
  $\mm$ mass regions (see text).
  The solid curve shows the result of the single-resonance fit described in
  the text; the dashed curve represents the background component.
}
\label{fg:fit-mY-jee_jmm}
\end{figure}

The $\pipiJ$ invariant mass spectrum for candidates passing all criteria 
is shown in Fig.~\ref{fg:fit-mY-jee_jmm} 
as points with error bars.  
Events that have an $e^+e^-$ ($\mu^+\mu^-$) mass in the $\jsi$ sidebands 
[2.76, 2.95] or [3.18, 3.25] ([2.93, 3.01] or [3.18, 3.25])\gevcc but 
pass all the other selection criteria are 
represented by the shaded histogram after being scaled 
by the ratio of the widths of the $\jsi$ mass window and sideband regions. 
An enhancement near 4.26\gevcc is clearly observed; no other structures 
are evident at the masses of 
the $J^{PC}\!=\!1^{--}$ charmonium states, $i.e.$, 
the $\psi(4040)$, $\psi(4160)$, and $\psi(4415)$~\cite{ct:PDG2004}, 
or the $\Xstat$. 
The Fig.~\ref{fg:fit-mY-jee_jmm} inset includes the $\psip$ region 
with a logarithmic scale for comparison;  
11802$\pm$110 $\psip$ events are observed, consistent with the 
expectation of 12142$\pm$809  $\psip$ events. 
We search for sources of backgrounds that contain a true $\jsi$ and 
peak in the $\pipiJ$ invariant mass spectrum.
The possibility that one or both pion candidates are misidentified kaons 
is checked by reconstructing the $\KK\jsi$ and $\KPi\jsi$ final states; 
we observe featureless mass spectra.
Similar studies of ISR events with a $\pipiJ$ plus one or more additional pions
reveal no structure that could feed down to produce a peak in the 
$\pipiJ$ mass spectrum. 
Two-photon events are studied directly by reversing 
the requirement on the missing mass; 
the number of events inferred for the signal region is a small fraction 
of those observed and their mass spectrum shows no structure. 
Hadronic $\ee\to\qqbar$ events produce $\jsi$ at a rate that 
is surprisingly large~\cite{ct:BaBar-prompt_ccbar, ct:Belle-prompt_ccbar, 
                            ct:Belle-double_ccbar, ct:BaBar-double_ccbar}, 
but no structure is observed for this background.

We evaluate the statistical significance of the enhancement using
unbinned maximum likelihood fits to the $\pipiJ$ mass spectrum. 
To evaluate the goodness of fit, the fit probability is determined 
from the $\chi^2$ and the number of degrees of freedom for bin
sizes of 5, 10, 20, 40, and 50\mevcc. Bins are combined with higher mass 
neighbors as needed to ensure that no bin is predicted to have 
fewer than seven entries. We try first-, second-, and third-order polynomials 
as null-hypothesis fit functions. 
The $\chi^2$-probability estimates for these fits range 
from $10^{-16}$ to $10^{-11}$. No substantial improvement
is obtained by including 
$\psi(4040)$,  $\psi(4160)$, or $\psi(4415)$~\cite{ct:PDG2004} 
terms in the fit. 
We conclude that the structure near 4.26\gevcc is statistically inconsistent 
with a polynomial background. 
Henceforth, we refer to this structure as the $\Ystat$.

\begin{figure}[t]
   \includegraphics[width=8cm]{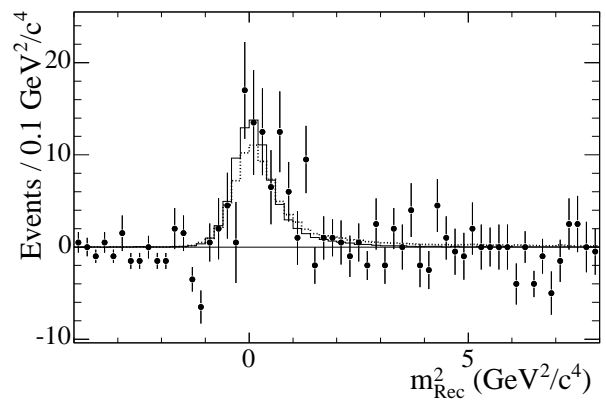}
\caption{The distribution of $\mrec^2$. 
         The points represent the data events
         passing all selection criteria except that on $\mrec^2$ and
         having a $\pipiJ$ mass near 4260\mevcc, minus the scaled
         distribution from neighboring $\pipiJ$ mass regions (see text).
         The solid histogram represents ISR $Y$ Monte Carlo events,
         and the dotted histogram represents 
         the ISR $\psip$ data events.}
\label{fg:Y-mRec2}
\end{figure}

It is important to test the ISR-production hypothesis 
because the $J^{PC}=1^{--}$ assignment for the $\Ystat$ follows from it. 
The ISR photon is reconstructed in (24$\pm$8)\% of the $\Ystat$ events, 
in agreement with the 25\% observed for ISR $\psip$ events. 
Kinematic distributions for the signal are obtained by subtracting
scaled distributions for events with $\pipiJ$ mass in the regions
[3.86, 4.06]\gevcc and [4.46, 4.66]\gevcc from those with
$\pipiJ$ mass in the signal region, defined as [4.16, 4.36]\gevcc.
The distribution of $\mrec^2$ is shown in Fig.~\ref{fg:Y-mRec2}, along with
corresponding distributions for ISR $\psip$ data events and for 
ISR $\Ystat$  Monte Carlo events.
Good agreement is found for these distributions, 
and for all other quantities studied 
to test that initial-state radiation is responsible for these events.

An unbinned likelihood fit to the  $\pipiJ$ mass spectrum 
is performed using a single relativistic Breit-Wigner signal function and 
a second-order polynomial background. 
The signal function is multiplied by a phase space factor and convoluted 
with the previously described resolution function. 
The fit gives 125$\pm$23 events 
with a mass of $4259\pm8\stat^{+2}_{-6}\syst$\mevcc 
and a width of $88\pm23\stat^{+6}_{-4}\syst$\mevcc. 
Systematic uncertainties include contributions from the fitting procedure, 
the mass scale, the mass-resolution function, and dependence on the
model of the $\Ystat \to \pipiJ$ decay. 
They have been added in quadrature. 
Under this single-resonance hypothesis we calculate a value of
$\Gamma(\Ystat\to\ee)\cdot\BR(\Ystat\to\pipiJ)=5.5\pm1.0^{+0.8}_{-0.7}$\evcc.
The fit probability determined from the $\chi^2$ and the number 
of degrees of freedom ranges from 0.3\% to 6.6\% for the same 
set of binning choices and background parameterizations 
used to evaluate the null hypothesis. 
To estimate the significance of the $\Ystat$ structure conservatively, 
we use instead of our optimized selection criteria, 
the criteria developed in analyzing just the first 124\invfb of data.  
Using these, we compare fits to the remaining 109\invfb of data sample 
with and without the resonance parameters determined by the first data sample. 
Using the binnings described above, we find a significance 
in the second independent data sample alone of 5 to 7$\sigma$. 
The likelihood and $\chi^2$ differences between signal and null-hypothesis 
fits to the full sample correspond to significances of at least 8$\sigma$. 

The robustness of the $\Ystat$ signal is 
tested with single-resonance fits to the $\pipiJ$ mass spectrum 
for $\ee$ and $\mm$ modes separately, which yield 
49$\pm$16 and 76$\pm$13 signal events, respectively. 
Fits give 76$\pm$18  events for the original
124\invfb data set and 56$\pm$13 events 
for the next, independent 109\invfb data set. 
Fitting samples with and without reconstructed ISR photons gives 30$\pm$11 
and 96$\pm$15 events, respectively.  
We find consistent values for the $\Ystat$ and the
$\psip$ when determining the fraction of the total signal found 
in each of these subsets.

Several additional systematic checks have been performed.
Each selection criterion has been tightened (loosened) and the
decrease (increase) in the signal yield is consistent with that for
the $\psip$ data.
Events selected when the selection criteria are reversed, 
individually or in pairs, are studied; 
in no case is there a significant dip in the signal-mass region that 
might indicate a bias in the selection procedure.

Since the single-resonance fit probability is low 
we consider the possibility 
that the observed signal is due to two interfering resonances. 
Two-resonance fits with an interference term find one resonance
mass close to the mass from the single-resonance fit, but with a
width as low as 50\mevcc, plus a second narrow resonance around
4.33\gevcc. 
However, the fit probabilities are not significantly improved
by two-resonance hypotheses. The size of our sample does not allow a 
statistically significant discrimination;
we can neither exclude nor establish a multi-resonance hypothesis.

\begin{figure}[t]
    \includegraphics[width=8cm]{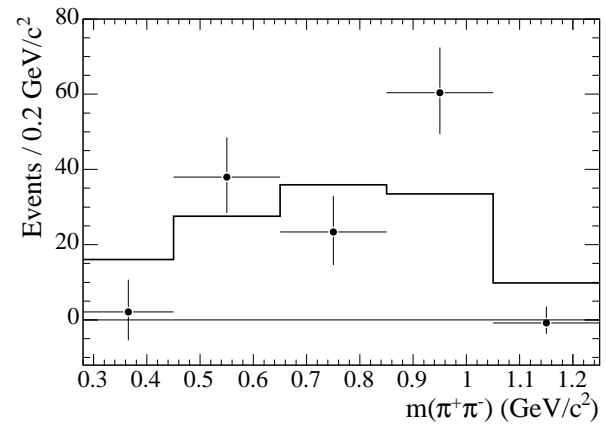}
\caption{The dipion mass distribution for $\Ystat\to\pipiJ$ data 
         is shown as points with error bars. 
         The histogram shows the distribution for Monte Carlo events 
         where $\Ystat\to\pipiJ$ is generated 
         according to an $S$-wave phase space model.}
\label{fg:m2Pi}
\end{figure}

The dipion invariant mass distribution for the $\Ystat$
is shown in Fig.~\ref{fg:m2Pi}. 
Each point represents the yield of a single-resonance
fit to the $\pipiJ$ mass distribution for that $\pi^+\pi^-$ mass bin.

No enhancement has been observed in the cross section for 
$\ee\to\textrm{hadrons}$~\cite{ct:PDG2004} at energies corresponding
to the $\Ystat$.
We compute the cross section for $\ee\to\pipiJ$ production 
at 4.25\gev, corresponding to the highest bin in our data,
to be about 50~pb. 
The inclusive hadronic cross section 
at $\sqrt{s}\!=\!4.25\gev$ is 14.2~nb~\cite{ct:PDG2004}.
The ratio, approximately 0.34\%, is smaller than the 4\% 
experimental uncertainty for the hadronic cross section, so this mode
would not have been visible. 
However, if the branching fraction of $\Ystat$ to $\pipiJ$ is very small, 
decays to other hadronic modes like $\DDbar$ would have been observable. 
This indicates that the branching fraction to $\pipiJ$ 
must be large compared to that for $\psi(3770)$~\cite{ct:BES-psi3770}. 

In summary, we have used initial-state radiation events to study the process
$\ee\to \pipiJ$ across the charmonium mass range.
In addition to the expected $\psip$ events, 
we observe an excess of 125$\pm$23 events centered 
at a mass of $\sim$4.26\gevcc,
signifying the presence of one or more previously unobserved
$J^{PC}=1^{--}$ states containing hidden charm.
At the current level of statistics we are unable to distinguish the 
number of new states; the data can be characterized by a single
resonance of mass $\sim$4.26\gevcc and of width $\sim$90\mevcc.

We are grateful for the excellent luminosity and machine conditions
provided by our \pep2\ colleagues, 
and for the substantial dedicated effort from
the computing organizations that support \babar.
The collaborating institutions wish to thank 
SLAC for its support and kind hospitality. 
This work is supported by
DOE
and NSF (USA),
NSERC (Canada),
IHEP (China),
CEA and
CNRS-IN2P3
(France),
BMBF and DFG
(Germany),
INFN (Italy),
FOM (The Netherlands),
NFR (Norway),
MIST (Russia), and
PPARC (United Kingdom). 
Individuals have received support from CONACyT (Mexico), A.~P.~Sloan Foundation, 
Research Corporation,
and Alexander von Humboldt Foundation.


\begin{thebibliography}{99}

  \bibitem{x-belle} 
     Belle Collaboration, S.~-K.~Choi {\it et al.}, 
     Phys.~Rev.~Lett. {\bf 91}, 262001 (2003).  

  \bibitem{cdf-B-X} CDF Collaboration, D.~Acosta {\it et al.}, 
     Phys.~Rev.~Lett. {\bf 93}, 072001 (2004).  

  \bibitem{d0-B-X} D0 Collaboration, V.~M.~Abazov {\it et al.}, 
     Phys.~Rev.~Lett. {\bf 93}, 162002 (2004).  

  \bibitem{babar-B-X} {\it \BaBar} Collaboration, B.~Aubert {\it et al.}, 
     Phys.~Rev.~D {\bf 71}, 071103(R) (2005).

  \bibitem{y3940-belle} 
     Belle Collaboration, S.~-K.~Choi {\it et al.}, 
     Phys.~Rev.~Lett. {\bf 94}, 182002 (2005).  

  \bibitem{ct:BaBar-ISR_X3872} {\it \BaBar} Collaboration, B.~Aubert {\it et al.}, 
     Phys.~Rev.~D {\bf 71}, 052001 (2005).
	

  \bibitem{babar-detector}
     \BaBar Collaboration, B.~Aubert {\it et al.}, 
     Nucl.~Instrum.~Methods Phys. Res., Sect.~A {\bf 479}, 1 (2002).


  \bibitem{lou-psi2s}  
     X.C.~Lou, Int.~J.~Mod.~Phys., Vol. 16, No. supp01B, 486 (2001).

  \bibitem{optimization} 
     G.~Punzi, 
     in {\it Proceedings of the Conference on Statistical Problems 
             in Particle Physics, Astrophysics and Cosmology (PHYSTAT2003), 
             Stanford, California, 2003,}
     eConf {\bf C030908} (2003) MODT002 [arXiv: physics/0308063].
     The value of 3/2 corresponds to an optimization for a signal of 3$\sigma$ significance.

  \bibitem{ct:Cauchy} 
    A non-relativistic Breit-Wigner shape.

  \bibitem{ct:PDG2004} Particle Data Group, S.~Eidelman {\it et al.}, 
     Phys.~Lett.~B {\bf 592}, 1 (2004).

  \bibitem{ct:BaBar-prompt_ccbar} 
    \BaBar Collaboration, B.~Aubert {\it et al.}, 
     Phys.~Rev.~Lett. {\bf 87}, 162002 (2001)

  \bibitem{ct:Belle-prompt_ccbar}
    Belle Collaboration, K.~Abe {\it et al.},
    Phys.~Rev.~Lett. {\bf 88}, 052001 (2002)

  \bibitem{ct:Belle-double_ccbar}  Belle Collaboration, K.~Abe {\it et al.},
     Phys.~Rev.~D {\bf 70}, 071102 (2004).


  \bibitem{ct:BaBar-double_ccbar}
     {\it \BaBar} Collaboration, B.~Aubert {\it et al.}, 
     Phys.~Rev.~D {\bf 72}, 031101 (2005).

     \bibitem{ct:BES-psi3770} BES Collaboration, J.~Z.~Bai {\it et al.},
     Phys.~Lett.~B {\bf 605}, 63 (2005).

 \end{thebibliography}
\end{document}